\documentclass[pdflatex,sn-vancouver-num]{sn-jnl}

\usepackage{graphicx}%
\usepackage{multirow}%
\usepackage{amsmath,amssymb,amsfonts}%
\usepackage{amsthm}%
\usepackage{mathrsfs}%
\usepackage[title]{appendix}%
\usepackage{xcolor}%
\usepackage{textcomp}%
\usepackage{manyfoot}%
\usepackage{booktabs}%
\usepackage{tabularx}%
\usepackage{algorithm}%
\usepackage{algorithmicx}%
\usepackage{algpseudocode}%
\usepackage{listings}%

\usepackage{array}

\theoremstyle{thmstyleone}%
\theoremstyle{thmstyletwo}%

\theoremstyle{thmstylethree}%

\begin{document}

\title[Learning to read residual images for single-pixel photography]{Learning to read residual images for single-pixel photography}


\author[1]{\fnm{Hongjun} \sur{An}}\email{an.hongjun@mail.nwpu.edu.cn}

\author[1]{\fnm{Shuaizhe} \sur{An}}\email{}

\author[3,4]{\fnm{Tong} \sur{Tian}}\email{}

\author*[1,2]{\fnm{Zhe} \sur{Sun}}\email{sunzhe@nwpu.edu.cn}

\affil[1]{\orgdiv{School of Artificial Intelligence, OPtics and ElectroNics (iOPEN)}, \orgname{Northwestern Polytechnical University}, \orgaddress{\city{Xi'an}, \postcode{710072}, \country{China}}}

\affil[2]{\orgdiv{Institute of Artificial Intelligence (TeleAI)}, \orgname{China Telecom}, \orgaddress{\city{Shanghai}, \postcode{200232}, \country{China}}}

\affil[3]{\orgdiv{Institute of Optics and Quantum Electronics, Abbe Center of Photonics}, \orgname{Friedrich Schiller University}, \orgaddress{\city{Jena}, \postcode{07743}, \country{Germany}}}

\affil[4]{\orgname{Helmholtz Institute Jena}, \orgaddress{\city{Jena}, \postcode{07743}, \country{Germany}}}


\abstract{
Continuous video scales in cost with camera count, resolution and recording duration, yet many monitoring tasks seek events rather than complete images and can operate on a compact single-pixel measurement stream.
Recovering a chosen interval as video is then desirable, but at 0.24\% of the pixel count the scene correlation image looks like noise and per-clip optimization takes minutes.
Correlating the residual of an estimate with the coding patterns nevertheless gives a residual image, the gradient of a centered measurement loss.
Lifted through a pretrained video decoder, its agreement with the true error exceeds that of control decoders.
Video-latent inversion (VLI) then outperforms frame-wise image-latent reconstruction at 1{,}000 bucket values.
Measurement-Gradient-Conditioned Video Flow (MGV-Flow), trained once to read the lifted residual image with preceding frames, reconstructs five frames in 1.55\,s at 22.9\,dB median PSNR, against 398\,s and 18.1\,dB for VLI, with the largest gains in high-motion scenes.
Directional fidelity ranks 30 coding configurations and selects one needing a third fewer bucket values.
On a physical system, MGV-Flow outperforms per-frame reconstruction without retraining.
A sliding window over the bucket-value stream yields 3.2 frames per second on one RTX~4090 and, in theory, 20 frames per second on a data-centre accelerator.
Demonstrations are available at \url{https://cavedweller509.github.io/MGV-Flow}.
}




\maketitle

\section{Introduction}\label{sec1}

\begin{figure*}[!tp]
\centering
\includegraphics[width=\textwidth]{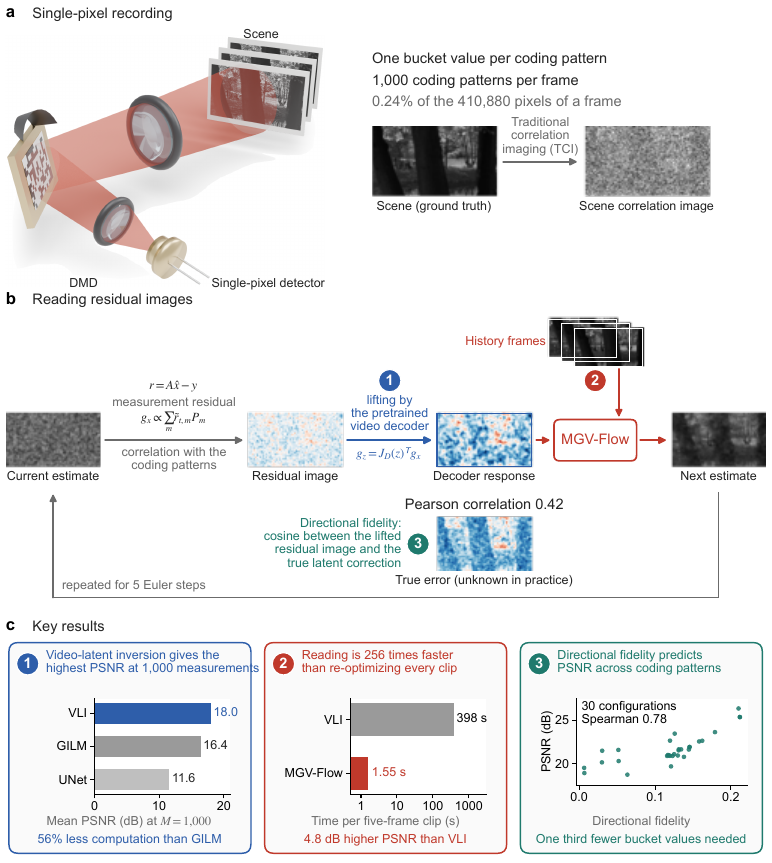}
\caption{
Reading residual images for single-pixel photography.
\textbf{a}, A digital micromirror device (DMD) displays binary coding patterns on the image of the scene, and a single-pixel detector records one bucket value per pattern. At 1{,}000 bucket values per frame, 0.24\% of the pixel count, traditional correlation imaging (TCI) gives a noise-like scene correlation image (rendered schematic).
\textbf{b}, The measurement residual $r$ between the bucket values $A\hat x$ predicted from the current estimate $\hat x$ by the forward operator $A$ and the acquired values $y$, centered within each frame (tilde) and correlated with the coding patterns $P_m$, is the residual image $g_x$. The transposed Jacobian of a pretrained video decoder $D$ lifts it to $g_z$, whose decoder response resembles the true error. Measurement-Gradient-Conditioned Video Flow (MGV-Flow) reads it with the history frames for five Euler steps. Badges 1 to 3 mark the three contributions.
\textbf{c}, Key results. (1) Peak signal-to-noise ratio (PSNR) of a pixel-space U-Net (UNet), frame-wise image-latent reconstruction (GILM) and video-latent inversion (VLI). (2) Time per five-frame clip. (3) Directional fidelity against PSNR for 30 coding configurations.
}
\label{fig:overview}
\end{figure*}

Recording the world as continuous video becomes costly at scale, because every additional camera, every gain in spatial resolution, and every additional hour of operation adds to the data that must be transmitted and stored.
Many monitoring and autonomous tasks, however, seek events, states, or anomalies rather than complete images and can operate directly on compressed or single-pixel optical measurements \cite{chen2017anomaly,ota2018ghost,fu2020sensing,xu2020compressed,cao2021classification,zhan2024global}, so a sensing system can keep a compact measurement stream during routine operation and recover a chosen interval of video only when a person needs to inspect it (Fig.~\ref{fig:overview}a).

Single-pixel imaging records such a stream by modulating the light from a scene with a sequence of known spatial patterns and integrating the modulated light on one detector into a single number per pattern, termed a bucket value \cite{takhar2006compressive,duarte2008single,edgar2019principles,gibson2020review}.
Compressive sensing, deterministic coding, and learned reconstruction have extended recovery under restricted budgets \cite{donoho2006compressed,katz2009compressive,zhang2015fourier,sun2017russian,lyu2017deep,wang2022physics,chen2025large}, and dynamic systems have carried single-pixel acquisition from static scenes to video \cite{edgar2015simultaneous,higham2018deep,lorente2020recurrent,xu2024hyperspectral,xu2026broadband}.
In the setting studied here, a five-frame clip at $856\times480$ pixels is recorded with 1{,}000 bucket values per frame, about 0.24\% of the pixel count, and at this rate correlation imaging, which weights each pattern with its mean-subtracted bucket value and sums the weighted patterns \cite{shapiro2008computational,ferri2010differential,sun2012normalized}, yields a map that looks like noise (Fig.~\ref{fig:residual-image}b, second column).
The scene must therefore be recovered from measurements that cannot reveal it, and quickly enough that reconstruction does not cancel the benefit of recording so little.

Although the measurements cannot show the scene, they can test any estimate of it.
For an estimated video $\hat x$, the known patterns define a linear forward operator $A$, and the measurement residual $r=A\hat x-y$ between the predicted and acquired bucket values is directly computable.
Under the noiseless model $y=Ax^\ast$ considered here, $r=-A(x^\ast-\hat x)$ is the projection into measurement space of the true error $x^\ast-\hat x$, which is the correction that the estimate needs.
Applying the same correlation to the residual instead of the bucket values carries this projection back to image space, and we call the result the residual image (Fig.~\ref{fig:residual-image}a).
With the normalization given in Methods, Section~\ref{sec4-measurement}, it is
\begin{equation}
(g_x)_t
=
\frac{1}{TMHW}
\sum_{m=1}^{M}\tilde r_{t,m}P_m,
\label{eq:rgi}
\end{equation}
where $(g_x)_t$ is the map for frame $t$, $T$ is the number of frames, $M$ the number of patterns per frame, $H$ and $W$ the frame dimensions, $P_m$ the $m$th pattern, and $\tilde r_{t,m}$ the residual of frame $t$ and pattern $m$ after subtraction of the mean residual of that frame.
For the linear model, $g_x$ is exactly the gradient of the mean-centered squared measurement residual with respect to the estimated frames, a relation that has driven iterative X-ray single-pixel tomography \cite{kingston2019ghosttomography} and links correlation with gradient descent in correlation imaging \cite{yu2021gradient}.
Like the scene correlation image, however, the residual image looks like noise (Fig.~\ref{fig:residual-image}b, third column).

As a weighted sum of the $M$ patterns, the residual image varies within each frame along at most $M$ of the $HW$ pixel directions, so the remaining components of the true error leave no trace in the bucket values and can only be supplied by prior knowledge of what natural video looks like.
Latent representations hold such knowledge in a compact variable and a decoder \cite{barbastathis2019deep,ongie2020deep,rombach2022latent}, and they have been used to impose single-pixel measurements on image latents in GILM \cite{chen2025large}, to couple optical acquisition with a learned latent space \cite{souza2025latent}, and to recover evolving physical fields from single-pixel data \cite{sholokhov2024single}.
For a video decoded as $\hat x=D(z)$ from a latent variable $z$, the chain rule lifts the residual image into the latent space as the lifted residual image $g_z=J_D(z)^{\mathsf T}g_x$, where $J_D(z)$ is the Jacobian of the decoder $D$ (Fig.~\ref{fig:residual-image}a).
The change that a step along $-g_z$ produces in the decoded video, the decoder response $J_D(z)(-g_z)$, is spatially structured where the residual image is not (Fig.~\ref{fig:residual-image}b, fourth column), and comparisons within spatial-frequency bands against decoders with re-initialized or shuffled weights show that this structure comes from the pretrained weights rather than from smoothing, with a mean band coherence with the true error of 0.076 against 0.029 for the residual image and at most 0.021 for the control decoders (Fig.~\ref{fig:residual-image}c).
The decoder response adds no information beyond the measurements and the current estimate, since it is the measurement-derived direction reshaped by the local Jacobian of the pretrained decoder, and its closer agreement with the true error indicates empirically that this reshaping favors the kinds of change that natural video exhibits in the components the measurements leave undetermined.
On this basis we formulate joint video-latent single-pixel reconstruction, which recovers the frames of a clip together rather than one by one, and refer to it as video-latent inversion (VLI).

Like other untrained and per-instance methods \cite{ulyanov2018deep,liu2021untrained,li2023urnet,li2024part,chen2025large}, VLI fits the measurements anew for every clip, and at about 80~s per frame a 100-frame interval would take more than two hours.
At 0.24\% sampling, many videos also agree with the measurements almost equally well, so a closer fit to them does not identify the scene.
Flow-matching and temporal models learn video dynamics from data and generate a video by integrating a learned velocity field from noise in a few steps \cite{lipman2023flow,liu2023flow,davtyan2023river,lorente2020recurrent,sholokhov2024single}, posterior-guidance methods insert the measurement gradient into diffusion or flow sampling as a prescribed correction at every step, also in latent spaces and for video \cite{chung2023dps,rout2023solving,song2024resample,benhamu2024dflow,martin2025pnpflow,kim2025flowdps,kwon2025solving}, diffusion priors have begun to enter single-pixel imaging \cite{song2024fspi}, amortized conditional flow models have been trained once to solve entire families of inverse problems without per-instance optimisation \cite{mammadov2024aps,dasgupta2026cfm}, and learned inverse solvers have shown that measurement gradients can serve as network inputs \cite{adler2017learned,putzky2017rim}.
We introduce Measurement-Gradient-Conditioned Video Flow (MGV-Flow, Fig.~\ref{fig:overview}b), a video-latent flow model that reads the lifted residual image together with the frames preceding the target interval, so that the preceding frames indicate what kind of video is plausible while the residual image indicates what was actually measured.
Because the model predicts the complete velocity instead of adding the feedback with a fixed weight, it is trained once and reconstructs each new clip in a few integration steps.

When a model reads a direction, the value of a coding pattern lies in how accurately the resulting feedback points toward the true latent correction.
Pattern families, scales, and acquisition orders have been chosen, adapted, and learned to improve image recovery \cite{zhang2015fourier,sun2017russian,yu2019cake,vaz2020hadamard,assmann2013adaptive,rousset2017adaptive,feng2020optimal,yuan2021adaptive,yan2023sampling}, and measurements have been selected by reducing speckle redundancy or by expected information gain \cite{chen2023turbulent,sun2026aime}, but these criteria concern the image rather than the feedback read in a latent representation.
We therefore propose directional fidelity, the cosine of the angle between $-g_z$ and the direction from the current latent to the latent of the ground-truth video (Fig.~\ref{fig:residual-image}d), and examine how pattern family, grain size (the width of one pattern cell), measurement count, and latent representation shape it and whether it predicts reconstruction quality.
Because it requires the ground truth, directional fidelity is an offline criterion for choosing coding configurations, and we test it by calibrating on some video clips and evaluating on disjoint ones.

Here we show that a single-pixel photography system can learn to read residual images (Fig.~\ref{fig:overview}c).
Joint inversion of a clip in the video latent space outperforms frame-wise image-latent reconstruction at low measurement counts and needs less computation and memory.
MGV-Flow reconstructs a five-frame clip in about a second instead of minutes and with higher quality than per-clip inversion, even though inversion fits the measurements more closely, and measurements of another video lower its quality, so the model uses the measured scene.
Run as a sliding window over the continuous stream of bucket values, it sustains 3.2~frames per second on one RTX~4090 with a reconstruction delay of 1.55~s, and in theory a single data-centre accelerator would reach 20~frames per second with a delay of about a quarter of a second.
Its average benefit over a flow model given only the preceding frames is small, however, and the measurements help mainly in high-motion scenes, where the preceding frames leave the future most uncertain.
Directional fidelity ranks coding configurations by the quality of physics-only steps without running it and selects a configuration that needs one third fewer measurements than a median-ranked one.
These analyses use simulated measurements, and on a physical single-pixel system with an accurate video history, MGV-Flow runs without retraining and outperforms per-frame reconstruction of the same frames.
Correlation imaging forms an image at normal sampling rates, but at extreme compression the same correlation applied to the residual returns instead as feedback that a generative model learns to read.



\section{Results}\label{sec2}

\subsection{Lifting residual images into a video latent space makes inversion scalable}

\begin{figure*}[!tp]
\centering
\includegraphics[width=\textwidth]{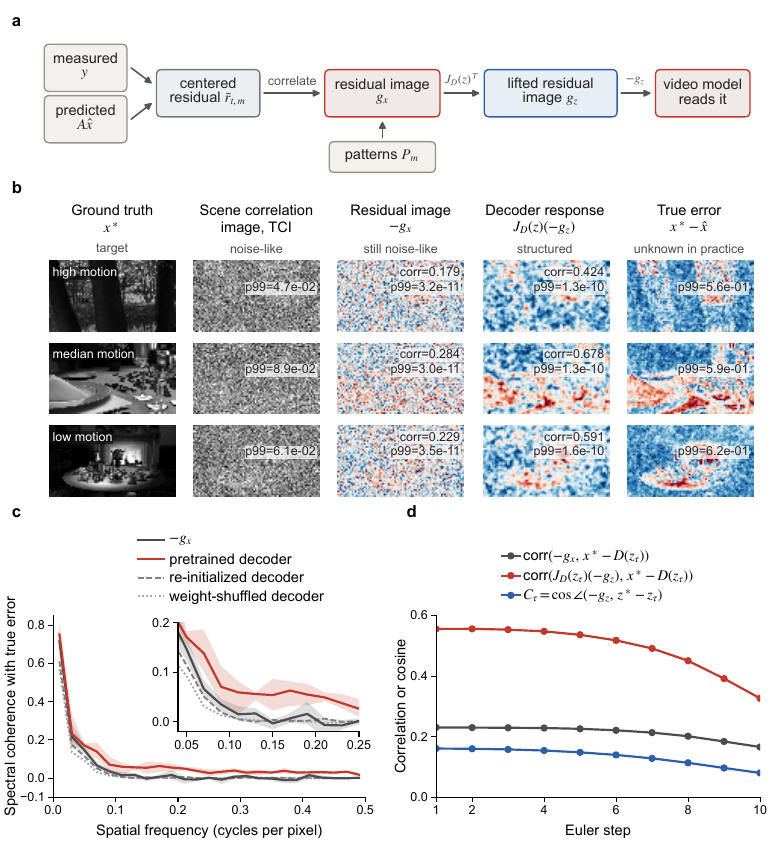}
\caption{
A noise-like residual image becomes a structured video correction through the pretrained decoder.
\textbf{a}, The residual between predicted ($A\hat x$) and acquired ($y$) bucket values, centered within each frame ($\tilde r_{t,m}$) and correlated with the coding patterns $P_m$, gives $g_x$ of equation~(\ref{eq:gx}), which the transposed Jacobian of the LeanVAE decoder $D$ at the latent $z$ lifts to $g_z$ of equation~(\ref{eq:gz}).
\textbf{b}, Ground truth $x^\ast$, scene correlation image of traditional correlation imaging (TCI), residual image $-g_x$, decoder response $J_D(z)(-g_z)$ and true error $x^\ast-\hat x$ for windows of highest, median and lowest motion (frame 3, first of $K=5$ Euler steps of Measurement-Gradient-Conditioned Video Flow). Corr is the Pearson correlation with the true error, and p99 the 99th percentile of absolute value that scales each tile.
\textbf{c}, Band coherence of equation~(\ref{eq:coherence}) with the true error for $-g_x$ and the decoder responses of pretrained, re-initialized and weight-shuffled decoders. Lines are means over 15 window-frame samples and three seeds for controls, bands interquartile ranges, and the inset enlarges 0.04 to 0.25 cycles per pixel.
\textbf{d}, Directional fidelity $C_\tau=\cos\angle(-g_z,z^\ast-z_\tau)$ along a $K=10$ trajectory with latent $z_\tau$ at flow time $\tau$ and ground-truth latent $z^\ast$, averaged over the 15 samples.
}
\label{fig:residual-image}
\end{figure*}

The reconstruction space determines where an iterative method computes the residual image, and we compared five methods that span three such spaces. Traditional correlation imaging (TCI) \cite{shapiro2008computational,gibson2020review} and two networks, a convolutional neural network (CNN) and a U-Net (UNet) \cite{lyu2017deep,he2018ghost,rizvi2020deepghost}, work in pixel space, GILM \cite{chen2025large} reconstructs each frame separately in the latent space of an image variational autoencoder (VAE), and video-latent inversion (VLI) optimizes a newly initialized video diffusion transformer \cite{peebles2023scalable} together with the pretrained decoder of the LeanVAE video autoencoder \cite{cheng2025leanvae} for every clip, so that five consecutive frames are reconstructed as one unit. Its comparison with GILM therefore reflects the video latent space and joint reconstruction together, whereas the frozen-decoder analysis of Fig.~\ref{fig:residual-image} isolates the pretrained weights. TCI forms the scene correlation image directly from the bucket values, whereas CNN, UNet, GILM and VLI minimize a per-frame standardized measurement loss by backpropagation through the operator, so that their update signal is a standardized relative of the residual image $g_x$ of equation~(\ref{eq:gx}), which GILM and VLI carry further through their decoders. The comparison covered four resolutions from $256\times144$ to $856\times480$ and $M$, the number of measurements per frame, from $1{,}000$ to $10{,}000$ across five test videos, with quality reported as peak signal-to-noise ratio (PSNR) and structural similarity (SSIM) \cite{wang2004image}.

\begin{figure*}[!tp]
\centering
\includegraphics[width=\textwidth]{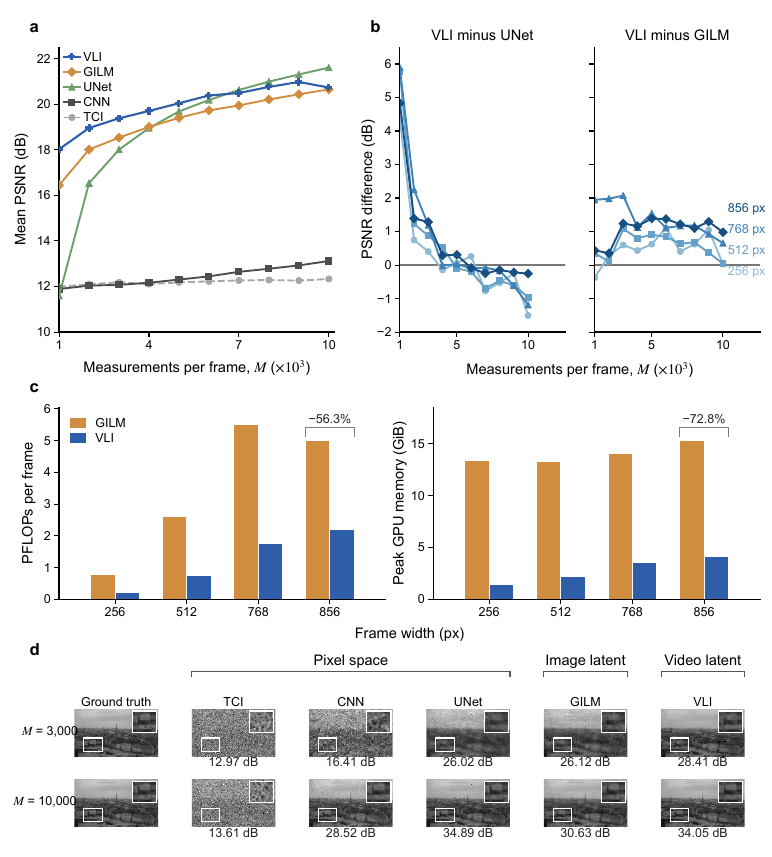}
\caption{
Measurement, spatial, and computational scaling across reconstruction spaces.
\textbf{a}, Peak signal-to-noise ratio, PSNR, in decibels, dB, versus $M$, the coding patterns or bucket values per frame from equation~(\ref{eq:forward}). The $\times10^3$ scale expresses $M$ in thousands. Methods are traditional correlation imaging, TCI, a convolutional neural network, CNN, a U-Net, UNet, GILM, the frame-wise image-latent method, and video-latent inversion, VLI.
\textbf{b}, PSNR advantage of VLI over UNet and GILM across $M$ and resolution.
\textbf{c}, PFLOPs are $10^{15}$ floating point operations per frame, GPU means graphics processing unit, and GiB means gibibyte. Percentages give reduction relative to GILM. The convergence time is defined by equation~(\ref{eq:t95}).
\textbf{d}, Reconstructions at $128\times72$ px, where px means pixels, for two values of $M$. Quality is reported as each run's best observed checkpoint averaged over four resolutions and five test videos.
}
\label{fig:latent-benchmark}
\end{figure*}

The advantage of the video latent is concentrated at low measurement counts (Fig.~\ref{fig:latent-benchmark} and Table~\ref{tab:quality-regime}). At $M=1{,}000$ and averaged over four resolutions and five test videos, VLI reaches a PSNR of 18.02~dB, compared with 16.44~dB for GILM and 11.59~dB for UNet, whereas once $M$ reaches $7{,}000$ and each frame is well constrained, UNet attains a higher PSNR while VLI retains the highest SSIM at every $M$. TCI remains near 12~dB across all $M$, because correlating raw bucket fluctuations with the patterns cannot form a usable image at these sampling ratios.

\begin{table*}[!tbp]
\caption{
Reconstruction quality across the evaluated optical measurement range.
Each entry reports mean peak signal-to-noise ratio, PSNR (dB), and structural similarity, SSIM, over four spatial resolutions and five test videos.
Bold values indicate the highest PSNR or SSIM at each measurement count.
For traditional correlation imaging (TCI), results at $M<10{,}000$ were recomputed using prefixes of the same $10{,}000$ measurement sequences.
}
\label{tab:quality-regime}

\centering
\small
\setlength{\tabcolsep}{4pt}

\begin{tabular}{@{}cccccc@{}}
\toprule
\shortstack{Measurements\\per frame\\$M$ (1)}
& TCI
& CNN
& UNet
& GILM
& VLI \\
& \shortstack{PSNR (dB)\\SSIM (1)}
& \shortstack{PSNR (dB)\\SSIM (1)}
& \shortstack{PSNR (dB)\\SSIM (1)}
& \shortstack{PSNR (dB)\\SSIM (1)}
& \shortstack{PSNR (dB)\\SSIM (1)} \\
\midrule

1,000
& 11.99 / 0.058
& 11.89 / 0.045
& 11.59 / 0.033
& 16.44 / 0.154
& \textbf{18.02} / \textbf{0.346} \\

2,000
& 12.10 / 0.062
& 12.04 / 0.051
& 16.53 / 0.163
& 18.01 / 0.276
& \textbf{18.96} / \textbf{0.376} \\

3,000
& 12.18 / 0.065
& 12.07 / 0.053
& 18.01 / 0.238
& 18.53 / 0.312
& \textbf{19.38} / \textbf{0.398} \\

4,000
& 12.10 / 0.066
& 12.16 / 0.056
& 18.95 / 0.305
& 19.00 / 0.325
& \textbf{19.70} / \textbf{0.423} \\

5,000
& 12.18 / 0.068
& 12.30 / 0.062
& 19.68 / 0.344
& 19.40 / 0.346
& \textbf{20.04} / \textbf{0.443} \\

6,000
& 12.22 / 0.069
& 12.44 / 0.069
& 20.19 / 0.373
& 19.73 / 0.365
& \textbf{20.38} / \textbf{0.464} \\

7,000
& 12.26 / 0.070
& 12.64 / 0.077
& \textbf{20.62} / 0.395
& 19.95 / 0.372
& 20.49 / \textbf{0.468} \\

8,000
& 12.28 / 0.072
& 12.78 / 0.083
& \textbf{20.99} / 0.413
& 20.21 / 0.381
& 20.76 / \textbf{0.486} \\

9,000
& 12.26 / 0.074
& 12.93 / 0.090
& \textbf{21.31} / 0.430
& 20.43 / 0.397
& 20.98 / \textbf{0.499} \\

10,000
& 12.33 / 0.076
& 13.12 / 0.100
& \textbf{21.61} / 0.443
& 20.65 / 0.400
& 20.73 / \textbf{0.491} \\

\bottomrule
\end{tabular}
\end{table*}

The gain over pixel networks is set mainly by the number of measurements per frame, whereas the advantage over a frame-wise image latent, averaged over all $M$ from the PSNR at the final update of each run, is 0.48 dB at the lowest resolution of $256\times144$ and between 1.05 and 1.37 dB at the two highest resolutions of $768\times432$ and $856\times480$ (Fig.~\ref{fig:latent-benchmark}b), consistent with a video latent designed for high resolution and few measurements per frame. At $M=1{,}000$, the gap over UNet stays between 4.37 and 5.89~dB at every resolution, although the measurement ratio falls from 2.71\% at $256\times144$ to 0.24\% at $856\times480$, while at $256\times144$ raising $M$ to $10{,}000$ changes this gap from 4.37 to $-1.50$~dB. Against GILM at $M=1{,}000$, the ordering reverses with resolution, from 18.32 versus 17.94~dB in favor of GILM at $256\times144$ to 15.40 versus 14.96~dB in favor of VLI at $856\times480$.

Joint video-latent inversion also reduces the cost of frame-wise image-latent inversion at high resolution (Fig.~\ref{fig:latent-benchmark}c and Table~\ref{tab:latent-efficiency}). At $856\times480$ and $M=10{,}000$, computation falls from 5.004 to 2.187~PFLOPs per frame, where one PFLOP is $10^{15}$ floating-point operations, a reduction of 56.3\%, and incremental peak graphics processing unit (GPU) memory falls from 15.31 to 4.17~GiB, a reduction of 72.8\%. The convergence time, the time needed to reach 95\% of the PSNR gain from equation~(\ref{eq:t95}), falls from 64.6 to 33.2~s per frame on an RTX~4090 and from 117.6 to 53.7~s per frame on an RTX~3090. The computation of GILM in Fig.~\ref{fig:latent-benchmark}c is lower at $856\times480$ than at $768\times432$ because computation is counted up to the best observed checkpoint, which GILM reached after a median of 300 and 400 of its 500 updates at these two resolutions.

\begin{table*}[!tbp]
\caption{
Computational cost of frame-wise image-latent and joint video-latent inversion at $856\times480$ and $M=10{,}000$.
FLOPs are normalized per reconstructed frame.
Peak memory denotes the incremental process-local graphics processing unit (GPU) allocation.
Conv.\ time is the convergence time on an RTX~3090 or RTX~4090 GPU, as medians of repeated runs on the first five frames of one test video.
Timing results were obtained from three repeated runs on an exclusively allocated GPU.
}
\label{tab:latent-efficiency}

\centering
\small
\setlength{\tabcolsep}{4pt}

\begin{tabular}{@{}lcccc@{}}
\toprule
Method
& \shortstack{Compute\\(PFLOPs frame$^{-1}$)}
& \shortstack{Peak memory\\(GiB)}
& \shortstack{Conv.\ time,\\3090 (s frame$^{-1}$)}
& \shortstack{Conv.\ time,\\4090 (s frame$^{-1}$)} \\
\midrule

GILM
& 5.004
& 15.31
& 117.60
& 64.60 \\

VLI
& 2.187
& 4.17
& 53.70
& 33.20 \\

\midrule

Reduction
& 56.3\%
& 72.8\%
& 54.4\%
& 48.6\% \\

\bottomrule
\end{tabular}
\end{table*}

This scalable inversion still consumes the residual signal through per-instance optimization, fitting a video network anew to every clip in a deep-image-prior style at a cost of tens of seconds per frame.

\subsection{Reading instead of consuming residual images enables amortized video reconstruction}

\begin{figure*}[!tp]
\centering
\includegraphics[width=\textwidth]{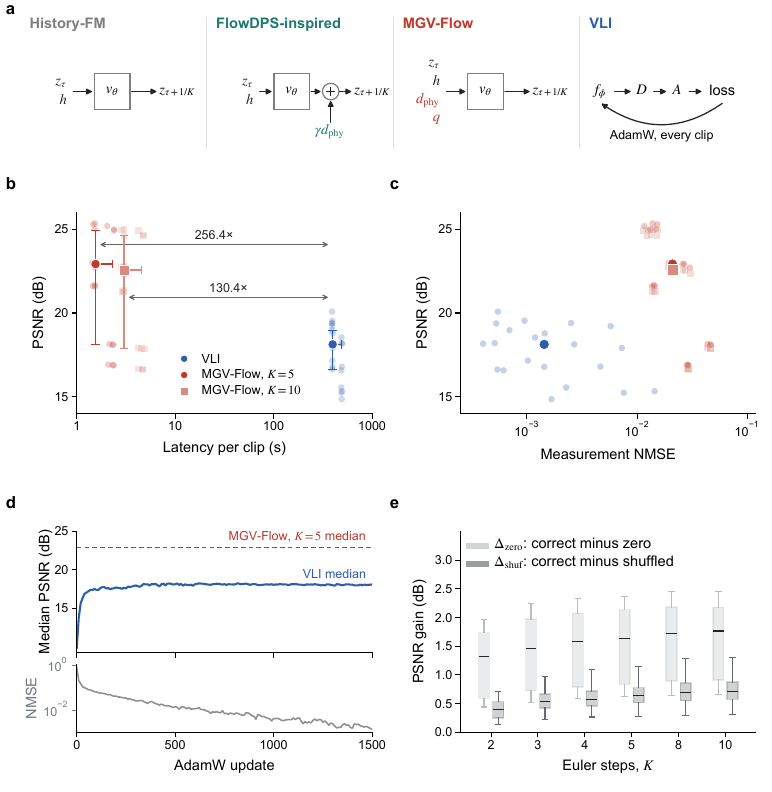}
\caption{
Reading versus consuming the residual image.
\textbf{a}, Symbols follow equations~(\ref{eq:residual}) to~(\ref{eq:flowdps}) and~(\ref{eq:lspi}). Measurement-Gradient-Conditioned Video Flow (MGV-Flow) predicts the velocity $v_\theta$ at flow time $\tau$ and latent $z_\tau$ by reading the history $h$, the normalized physical direction $d_{\mathrm{phy}}$ and the quality descriptor $q$, and one Euler step gives the next latent $z_{\tau+1/K}$. The network $f_\phi$ takes a clip's encoded five-frame history $h$ as input, and video-latent inversion (VLI) optimizes $\phi$ and the pretrained decoder $D$ for every clip by backpropagating the single-pixel imaging (SPI) loss through the operator $A$ with AdamW, Adam with decoupled weight decay. RMS is root mean square. History-conditioned flow matching (History-FM) omits $d_{\mathrm{phy}}$ and $q$, and FlowDPS-inspired adds $\gamma d_{\mathrm{phy}}$ to the velocity.
\textbf{b}, Latency is in seconds, s, peak signal-to-noise ratio, PSNR, in decibels, dB, and $\times$ is speedup.
\textbf{c}, Median measurement normalized mean squared error, NMSE, from equation~(\ref{eq:nmse}) and PSNR are shown for $K=5$, the Euler step count.
\textbf{d}, $\Delta_{\mathrm{zero}}$ and $\Delta_{\mathrm{shuf}}$ are correct-conditioning PSNR minus zero- or shuffled-conditioning PSNR. The scope is 21 latency cases and 105 conditioning cases per $K$.
}
\label{fig:read-consume}
\end{figure*}

MGV-Flow is trained once to read the lifted residual image $g_z$ of equation~(\ref{eq:gz}), so that a new video segment requires a few integration steps instead of a fresh optimization. Its velocity $v_\theta$ reads the encoded five-frame history $h$, the normalized physical direction $d_{\mathrm{phy}}$ of equation~(\ref{eq:dphy}), and the five-dimensional quality descriptor $q$ of equation~(\ref{eq:q}), and the network decides how to use these inputs rather than adding them to its output (Fig.~\ref{fig:read-consume}a). We compared it with History-FM, a history-conditioned flow-matching (FM) model of the same architecture that receives only the history, with a FlowDPS-inspired explicit posterior-gradient baseline \cite{kim2025flowdps} that adds $\gamma d_{\mathrm{phy}}$ with fixed weight $\gamma=1.0$ at every step as in equation~(\ref{eq:flowdps}), and with VLI, which shares the LeanVAE latent space but optimizes its network anew for every clip. Each test case is a window of ten consecutive frames, five history frames followed by five target frames, measured with $M=1{,}000$ at $856\times480$ and reconstructed with $K$ Euler steps of equation~(\ref{eq:euler}).

A five-frame clip took 397.99~s with VLI and 1.55~s with MGV-Flow at $K=5$, a paired median speedup of 256.4$\times$ with all methods of a case on the same RTX~4090, while median PSNR over the final results of 21 validation cases rose from 18.12 to 22.92~dB and SSIM from 0.323 to 0.713 (Fig.~\ref{fig:read-consume}b and Table~\ref{tab:mgv-latency}). With $K=10$, MGV-Flow took 3.04~s and reached 22.56~dB.

\begin{table*}[!tbp]
\caption{
Matched reconstruction latency at $856\times480$ and $M=1{,}000$.
Latency and reconstruction metrics are medians over 21 paired cases after taking the median over three timing repeats within each case.
All methods within a paired case used identical bucket values and were executed on the same exclusive RTX~4090 graphics processing unit (GPU).
Speedup is the median of the case-wise end-to-end (E2E) latency ratio relative to VLI, video-latent inversion. PSNR is the peak signal-to-noise ratio, SSIM the structural similarity, and NMSE the normalized mean squared error.
}
\label{tab:mgv-latency}
\centering
\small
\setlength{\tabcolsep}{4pt}
\begin{tabular}{@{}lcccccc@{}}
\toprule
Method
& \shortstack{E2E latency\\(s clip$^{-1}$)}
& \shortstack{E2E latency\\(s frame$^{-1}$)}
& \shortstack{PSNR\\(dB)}
& \shortstack{SSIM\\(1)}
& \shortstack{Measurement\\NMSE (1)}
& \shortstack{Speedup\\($\times$)} \\
\midrule
\shortstack[l]{VLI\\1,500 updates} & 397.99 & 79.60 & 18.12 & 0.323 & 0.00143 & 1.0 \\
MGV-Flow, $K=5$              & 1.55   & 0.31  & 22.92 & 0.713 & 0.0210  & 256.4 \\
MGV-Flow, $K=10$             & 3.04   & 0.61  & 22.56 & 0.702 & 0.0211  & 130.4 \\
\bottomrule
\end{tabular}
\end{table*}

The gain does not come from a closer fit to the measurements, because VLI fits them far better, with a median measurement normalized mean squared error (NMSE), defined in equation~(\ref{eq:nmse}), of 0.00143 against 0.0210 for MGV-Flow, and within 1,500 updates it matched the measurement fit of MGV-Flow in all 21 cases but its PSNR in none (Fig.~\ref{fig:read-consume}c). Because many videos fit the measurements almost equally well at 0.24\% sampling, a tighter fit does not identify the scene, whereas MGV-Flow reads the residual image together with the history and lets the learned prior choose, among the videos that fit, one that is plausible as video.

At $K=5$, PSNR was 21.86~dB with the correct measurements, 20.29~dB with the measurement-derived inputs set to zero, and 21.18~dB with inputs computed from the measurements of another video, giving $\Delta_{\mathrm{zero}}=1.57$~dB and $\Delta_{\mathrm{shuf}}=0.68$~dB (Fig.~\ref{fig:read-consume}d and Table~\ref{tab:mgv-conditioning}), and at $K=10$ the two gains grew to 1.65 and 0.76~dB, with measurement NMSE moving the same way at every $K$. Because the shuffled control supplies an input of the same kind and size, the gain is not explained by the extra input itself.

\begin{table*}[!tbp]
\caption{
Effect of target-consistent residual image conditioning over the validation cases.
Peak signal-to-noise ratio (PSNR) values are means over 105 matched reconstruction cases at each integration budget.
$\Delta_{\mathrm{zero}}$ and $\Delta_{\mathrm{shuf}}$ denote the PSNR difference of correct conditioning relative to zero and cross-video shuffled physics, respectively.
Measurement normalized mean squared error (NMSE) compares the reconstructed video with the target bucket values.
}
\label{tab:mgv-conditioning}
\centering
\small
\setlength{\tabcolsep}{4pt}
\begin{tabular}{@{}ccccccc@{}}
\toprule
\shortstack{Euler steps\\$K$ (1)}
& \shortstack{Correct PSNR\\(dB)}
& \shortstack{$\Delta_{\mathrm{zero}}$\\(dB)}
& \shortstack{$\Delta_{\mathrm{shuf}}$\\(dB)}
& \shortstack{Correct NMSE\\(1)}
& \shortstack{Zero NMSE\\(1)}
& \shortstack{Shuffled NMSE\\(1)} \\
\midrule
2  & 22.22 & +1.24 & +0.41 & 0.0328 & 0.0567 & 0.0484 \\
3  & 22.10 & +1.43 & +0.58 & 0.0303 & 0.0605 & 0.0519 \\
4  & 21.96 & +1.52 & +0.63 & 0.0295 & 0.0632 & 0.0541 \\
5  & 21.86 & +1.57 & +0.68 & 0.0291 & 0.0651 & 0.0561 \\
8  & 21.66 & +1.63 & +0.74 & 0.0287 & 0.0686 & 0.0596 \\
10 & 21.58 & +1.65 & +0.76 & 0.0287 & 0.0700 & 0.0610 \\
\bottomrule
\end{tabular}
\end{table*}

\begin{figure*}[!tp]
\centering
\includegraphics[width=\textwidth]{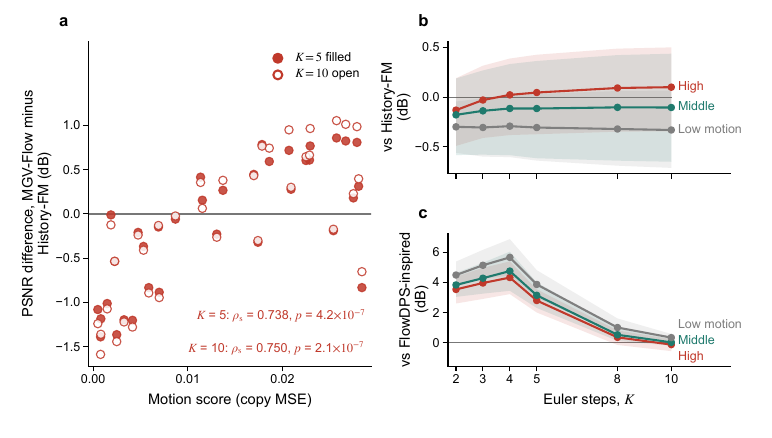}
\caption{
When reading the residual image pays off.
\textbf{a}, Motion score is the copy mean squared error, MSE, from the last history frame to five future frames. It is compared with the MGV-Flow PSNR difference versus History-FM, where MGV-Flow is Measurement-Gradient-Conditioned Video Flow, History-FM is history-conditioned flow matching, PSNR is peak signal-to-noise ratio in decibels, dB, and versus means that the named baseline is subtracted from MGV-Flow. Filled and open markers encode the displayed Euler step counts $K$. Annotations give the Spearman correlation $\rho_{\mathrm s}$ and its two-sided $p$ value.
\textbf{b}, Mean PSNR difference of MGV-Flow versus History-FM in the high, middle, and low motion groups, with the grouping rule fixed in advance, across $K$. Bands are 95\% bootstrap intervals over windows.
\textbf{c}, The same for MGV-Flow versus FlowDPS-inspired, the explicit posterior-gradient flow baseline. The scope of \textbf{b} and \textbf{c} is 35 validation windows with three seeds each.
}
\label{fig:motion}
\end{figure*}

Reading the measurements pays off when the history cannot predict the target. We ranked the five windows of each video by a motion score, the mean squared difference between the last history frame and the five target frames, with the grouping rule fixed in advance. The benefit of reading over History-FM grew with motion, with Spearman rank correlations of $\rho_{\mathrm s}=0.738$ at $K=5$ and 0.750 at $K=10$ across 35 windows ($p<10^{-6}$, Fig.~\ref{fig:motion}a). In the high-motion group it rose from $-0.131$~dB at $K=2$ to $+0.101$~dB at $K=10$, whereas in the low-motion group it stayed negative at $-0.330$~dB for $K=10$ (Fig.~\ref{fig:motion}b), so that averaged over all 105 cases MGV-Flow and History-FM lie within 0.15~dB of each other, at 21.86 and 21.99~dB for $K=5$. Adding the same direction to the History-FM velocity with the fixed weight $\gamma$, in contrast, lowered the mean PSNR to 18.56~dB at $K=5$, so that MGV-Flow exceeded FlowDPS-inspired by 2.80, 3.15, and 3.87~dB in the high, middle, and low motion groups (Fig.~\ref{fig:motion}c). The gap was largest where the history alone already predicted the target ($\rho_{\mathrm s}=-0.852$ at $K=5$) and closed only as the steps became smaller, to between $-0.13$ and $+0.33$~dB at $K=10$, where FlowDPS-inspired reached 21.49~dB, which is consistent with a fixed-weight correction that overshoots with coarse steps, whereas MGV-Flow learned during training how strongly to follow the direction.

On Something-Something V2 (SSv2) \cite{goyal2017something}, a dataset not used for training, the ranking depends on the number of steps. At $K=5$, MGV-Flow reached 19.67~dB against 19.98~dB for History-FM and 18.76~dB for FlowDPS-inspired and led in five of eight scenes, whereas at $K=10$ FlowDPS-inspired led with 21.59~dB against 19.71 and 19.45~dB and in six scenes. Explicit correction therefore became competitive only with more steps, where on unseen data it exploited the measurements more strongly than the learned reading, whereas MGV-Flow remained the more accurate measurement-based model with few steps.

\begin{figure*}[!tp]
\centering
\includegraphics[width=\textwidth,height=0.74\textheight,keepaspectratio]{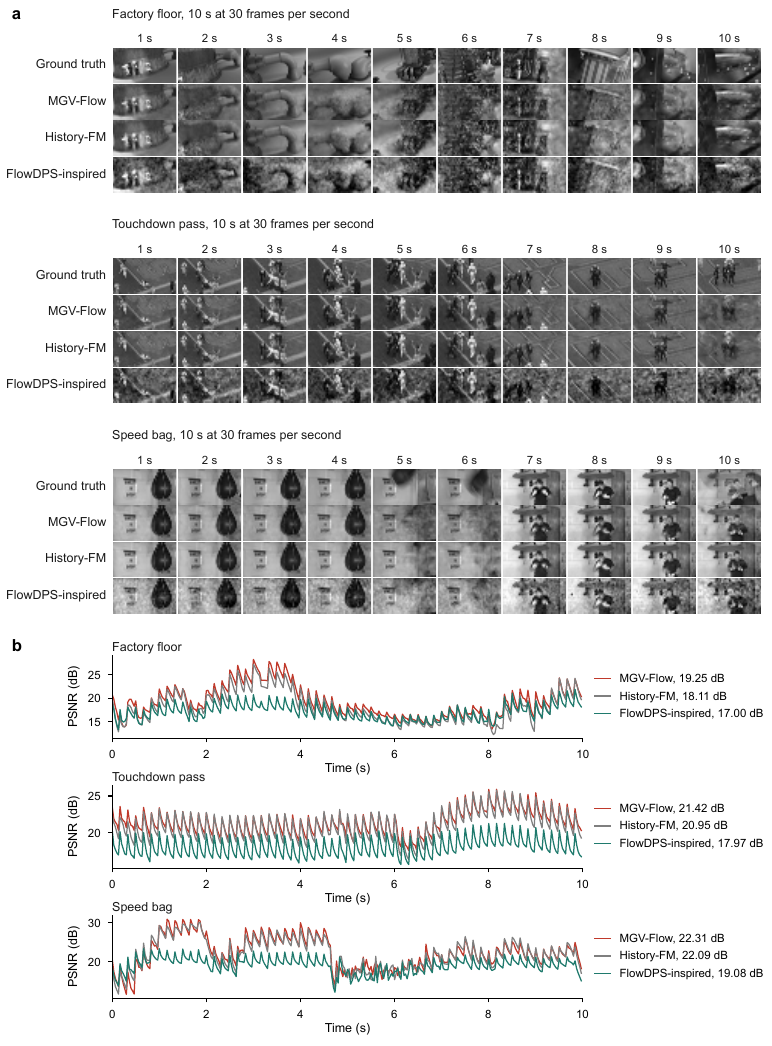}
\caption{
Rolling single-pixel photography of three 10~s high-motion clips.
\textbf{a}, Every block of five target frames is reconstructed from its own $M=1{,}000$ bucket values per frame with $K=5$ Euler steps, taking the five preceding ground-truth frames as history, and one reconstructed frame is shown per second. Measurement-Gradient-Conditioned Video Flow (MGV-Flow) is compared with history-conditioned flow matching (History-FM) and with FlowDPS-inspired, the explicit posterior-gradient flow baseline.
\textbf{b}, Peak signal-to-noise ratio (PSNR) in decibels (dB) of every reconstructed frame, with the clip mean in the legend. The periodic decline follows the position of a frame within its five-frame block.
}
\label{fig:video}
\end{figure*}

Each block of five target frames was measured with its own bucket values and reconstructed with $K=5$ from the five ground-truth frames that precede it, and the blocks were joined in order (Fig.~\ref{fig:video}). Across three high-motion clips with diverse scene content, MGV-Flow exceeded History-FM and FlowDPS-inspired on every clip, with mean PSNR of 19.25, 21.42, and 22.31~dB against 18.11, 20.95, and 22.09~dB for History-FM and 17.00, 17.97, and 19.08~dB for FlowDPS-inspired. FlowDPS-inspired frames show visible blotchy artefacts at this step count, and the gap to History-FM is largest on the factory clip, where motion is highest, which is consistent with the window-level finding that reading the measurements helps most where the history predicts least.

\subsection{Directional fidelity of residual images reveals coding-pattern utility}

The directional fidelity $C$ quantifies whether the lifted residual image points toward the true latent correction in video latent space. At the probe latent $z_{\mathrm{pr}}$ formed at flow time $\tau=0.5$ by equation~(\ref{eq:probe}), equation~(\ref{eq:directional-fidelity}) compares $-g_z(z_{\mathrm{pr}})$ with $z^\ast-z_{\mathrm{pr}}$, at the cost of one residual image and one decoder vector-Jacobian product and without running a reconstruction. We evaluated 30 combinations of pattern family, grain size, which is the width of one pattern cell in pixels (px), and measurement count on eight held-out windows, and compared $C$ with the terminal PSNR of the ten-step physics-only latent trajectory of equation~(\ref{eq:pure-physics}) from the same $z_{\mathrm{pr}}$. MGV-Flow was trained with one pattern family and one grain size and was therefore not part of this comparison, and whether the selected patterns also improve it remains to be tested by retraining.

\begin{figure*}[!tp]
\centering
\includegraphics[width=\textwidth]{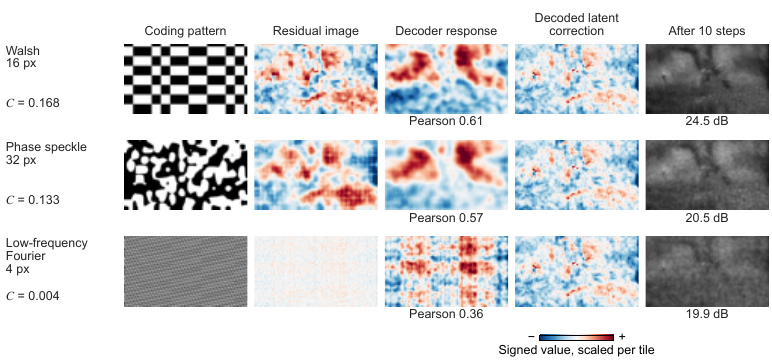}
\caption{
Residual images produced by three coding configurations at $M=3{,}000$ on one held-out window. Rows are the calibration-selected Walsh basis at 16 px, the median-ranked phase-speckle patterns at 32 px, and the low-frequency Fourier basis at 4 px, which has the lowest held-out mean directional fidelity of the 30 configurations. Columns show the first coding pattern, the residual image $-g_x$ at the probe latent $z_{\mathrm{pr}}$ of equation~(\ref{eq:probe}), the decoder response $J_D(z_{\mathrm{pr}})(-g_z)$, the decoded latent correction $D(z^\ast)-D(z_{\mathrm{pr}})$, which is the same in every row, and the reconstruction after the ten physics-only steps of equation~(\ref{eq:pure-physics}). $C$ is the directional fidelity of equation~(\ref{eq:directional-fidelity}), Pearson is the correlation coefficient between decoder response and decoded latent correction, and PSNR is the peak signal-to-noise ratio of the reconstruction, all averaged over the five frames of the clip. Images show frame 3. Signed maps use a symmetric scale clipped at the 99th percentile of absolute value of each tile. The window is the lower median of the Walsh 16 px fidelity among the eight held-out windows.
}
\label{fig:pattern-gradients}
\end{figure*}

For one held-out window and one starting latent, we compared three configurations (Fig.~\ref{fig:pattern-gradients}) at $M=3{,}000$, namely the calibration-selected Walsh basis at 16 px, the median-ranked phase speckle at 32 px, and the low-frequency Fourier basis at 4 px, which had the lowest held-out mean $C$ of all 30. The Walsh and phase-speckle residual images are blocky at the scale of their pattern cells, whereas the 4 px Fourier map is fine-grained noise. After lifting, the Walsh decoder response agrees best with the decoded latent correction, with a mean Pearson correlation over the five frames of 0.61 against 0.57 and 0.36, its directional fidelity is the highest at 0.168 against 0.133 and 0.004, and ten physics-only steps from the same start reach 24.5, 20.5, and 19.9~dB, so the pattern whose feedback points best also reconstructs best.

\begin{figure*}[!tp]
\centering
\includegraphics[width=\textwidth]{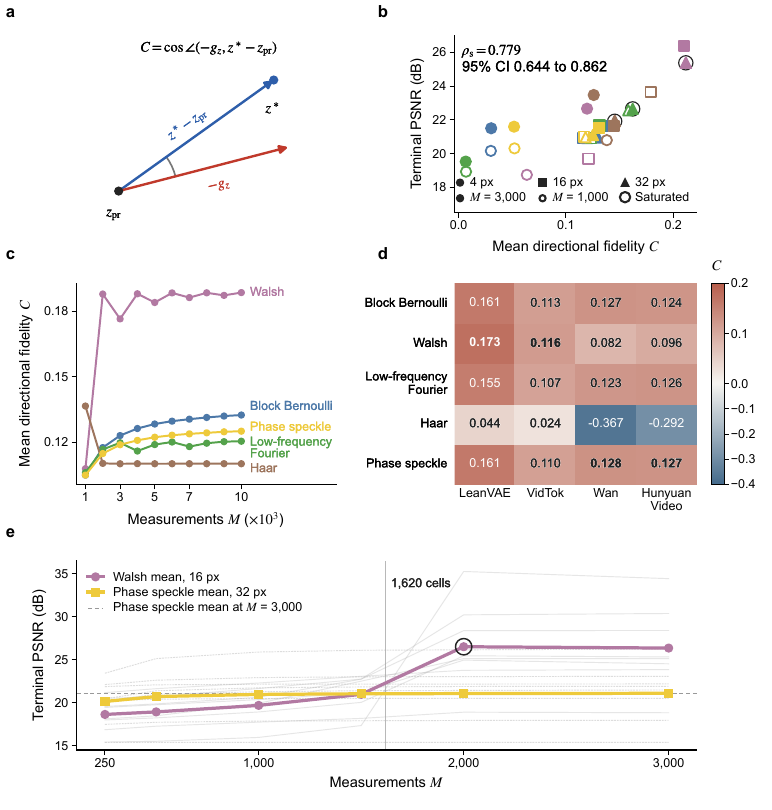}
\caption{
Directional fidelity reveals coding-pattern utility.
\textbf{a}, Directional fidelity is $C=\cos\angle(-g_z(z_{\mathrm{pr}}),z^\ast-z_{\mathrm{pr}})$, where $g_z$ is the lifted residual image from equation~(\ref{eq:gz}), $z_{\mathrm{pr}}$ is the probe latent at $\tau=0.5$ from equation~(\ref{eq:probe}), and $z^\ast=E(x^\ast)$ is the ground-truth video latent.
\textbf{b}, PSNR is peak signal-to-noise ratio in decibels, dB, $\rho_{\mathrm s}$ is Spearman correlation, CI is confidence interval, $M$ is patterns per frame, $\times10^3$ means thousands, and px means pixels. Saturated sets have exhausted their unique centered span at the shown $M$.
\textbf{c}, LeanVAE, the Lean video variational autoencoder, is shown for Block Bernoulli, Walsh, Low-frequency Fourier, Haar, and Phase speckle patterns.
\textbf{d}, These families are compared across LeanVAE, VidTok, Wan, and HunyuanVideo video autoencoder backends. VAE means variational autoencoder.
\textbf{e}, First match is the smallest tested $M$ where selected Walsh reaches the threshold. Coarse cells are spatial cells at the shown grain size. The scope is 30 configurations and eight held-out windows.
}
\label{fig:pattern-utility}
\end{figure*}

Directional fidelity depends jointly on the coding pattern, its grain, and the video representation (Fig.~\ref{fig:pattern-utility}b to d). With LeanVAE at a grain size of 16 px, the five families followed distinct responses as $M$ increased, and we repeated the analysis with three other video autoencoders, VidTok \cite{tang2024vidtok}, Wan \cite{wan2025wan}, and HunyuanVideo \cite{kong2024hunyuanvideo}. At $M=3{,}000$, Walsh ranked first for LeanVAE and VidTok, whereas for Wan and HunyuanVideo block Bernoulli, low-frequency Fourier, and phase speckle were nearly tied within 0.005 in mean $C$ and Walsh fell to the lowest positive value. Haar was near zero for LeanVAE and VidTok, with means of 0.044 and 0.024, and strongly negative for Wan and HunyuanVideo, with means of $-0.367$ and $-0.292$. The value of a coding pattern therefore depends on the latent representation.

Directional fidelity predicted end-to-end reconstruction quality across the held-out configurations (Fig.~\ref{fig:pattern-utility}b), with a Spearman rank correlation of $\rho_{\mathrm s}=0.779$ and a 95\% bootstrap confidence interval from 0.644 to 0.862 in the primary analysis, which was specified before the held-out results were examined. In analyses not specified in advance, the correlation stayed positive within each measurement count ($\rho_{\mathrm s}=0.764$ at $M=1{,}000$ and 0.654 at $M=3{,}000$), within each grain size (0.527, 0.855, and 0.939 at 4, 16, and 32 px), and after removing the three 32 px bases whose distinct patterns run out before $M=3{,}000$ (0.726 over 27 configurations). Within families it was strong for low-frequency Fourier (0.943) and Walsh (0.771), moderate for block Bernoulli (0.657) and Haar (0.429), and absent for phase speckle (0.086), so directional fidelity discriminates best between pattern families and between structured patterns of different scale.

We ranked the 30 configurations by their mean $C$ on four calibration windows and evaluated this ranking on the eight held-out windows (Fig.~\ref{fig:pattern-utility}e). Calibration rank 1, Walsh at 16 px and $M=3{,}000$, reached 26.37~dB, which is 5.27~dB above the median-ranked configuration at calibration rank 15, phase speckle at 32 px and $M=3{,}000$ with 21.10~dB, and 4.69~dB above uniform random selection. In a test that was also specified in advance, the selected configuration rose from 20.98~dB at $M=1{,}500$ to 26.52~dB at $M=2{,}000$, the smallest tested count that matched the median-ranked configuration at $M=3{,}000$. This step coincides with the measurements becoming complete at the 16 px coding scale, whose $30\times54$ coarse grid contains 1{,}620 cells. The rank of the 16 px Walsh basis restricted to this coarse grid is 1{,}270 at $M=1{,}500$ and full, 1{,}619 of 1{,}620 cells after mean removal, at $M=2{,}000$, whereas the 32 px phase-speckle comparator levels off near 21~dB. At $M=2{,}000$, the selected configuration exceeded the median-ranked configuration at $M=3{,}000$ in all eight held-out windows, with a median per-window gain of 3.60~dB and a minimum of 0.70~dB, so directional fidelity identified a configuration that reached the comparator quality with one third fewer measurements.

\subsection{On physical measurements, reading transfers without retraining}\label{sec2-physical}

\begin{figure*}[!tbp]
\centering
\includegraphics[width=\textwidth]{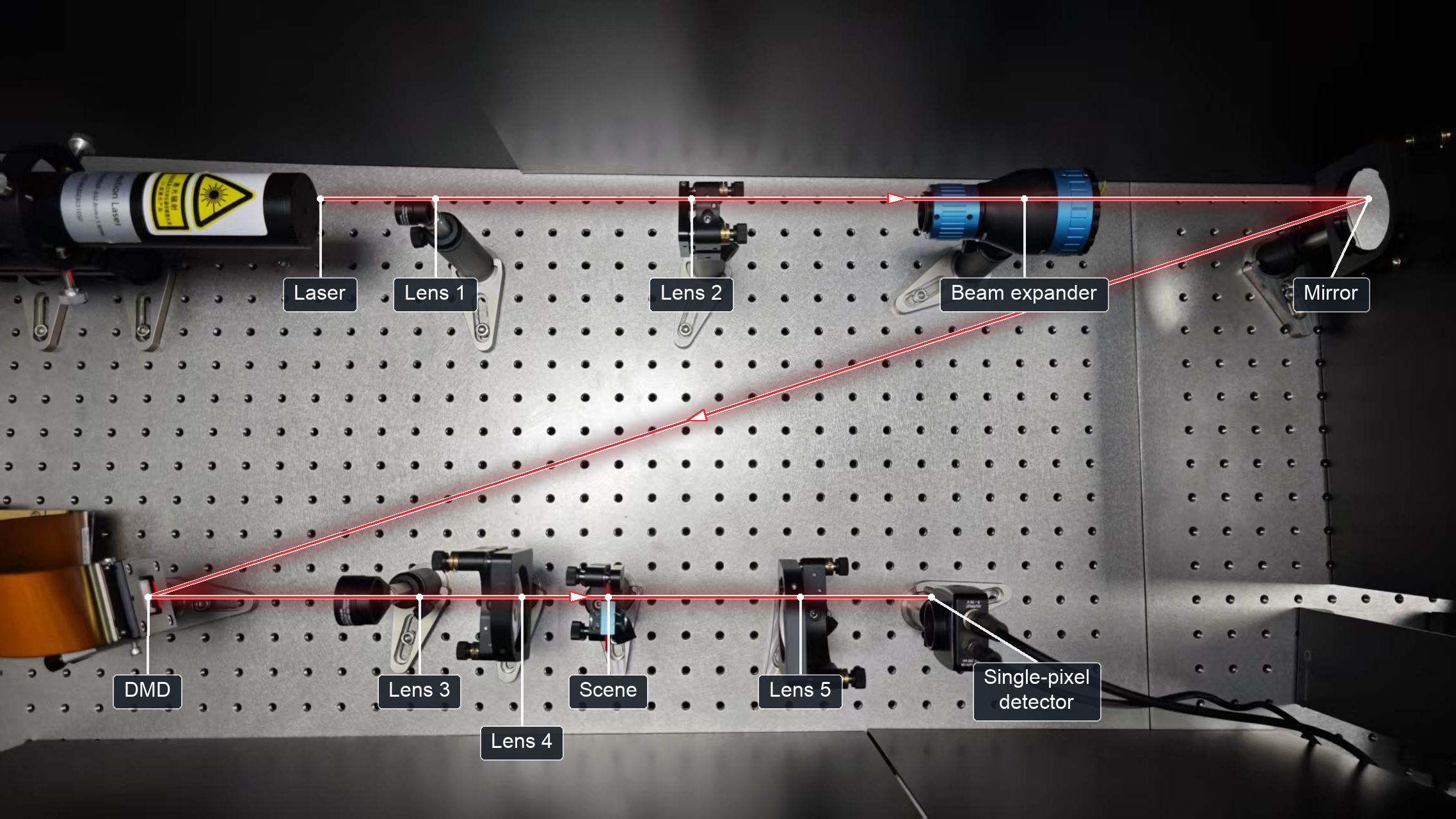}
\caption{
Physical single-pixel photography system, top view. A laser beam passes two lenses and a beam expander, is folded by a mirror onto a digital micromirror device (DMD) that displays the binary coding patterns, and is projected by two lenses onto the scene. A fifth lens collects the light from the scene onto the single-pixel detector, which records one bucket value per pattern. The red line marks the beam path.
}
\label{fig:optical-setup}
\end{figure*}

We tested reading on bucket values acquired by a physical single-pixel photography system with structured illumination (Fig.~\ref{fig:optical-setup}). Two ten-frame excerpts of the animated films Plane Crazy and Steamboat Willie were acquired frame by frame with the same ordered bank of 10{,}000 binary patterns that the simulation uses, about 2.4\% of the $856\times480$ pixel count. MGV-Flow, neither retrained nor fine-tuned on physical data, predicted frames 6 to 10 of each clip from their measured bucket values with the ground-truth frames 1 to 5 as history. Estimates and references were rectified to the illuminated region, and because the detector response differs photometrically from the simulated operator, we report PSNR and SSIM before and after independent histogram equalization, using the equalized values as the primary measure (Methods, Section~\ref{sec4-physical}). As references, CNN, UNet, and GILM reconstructed each of the same frames independently from its own bucket values.

\begin{figure*}[!tp]
\centering
\includegraphics[width=\textwidth]{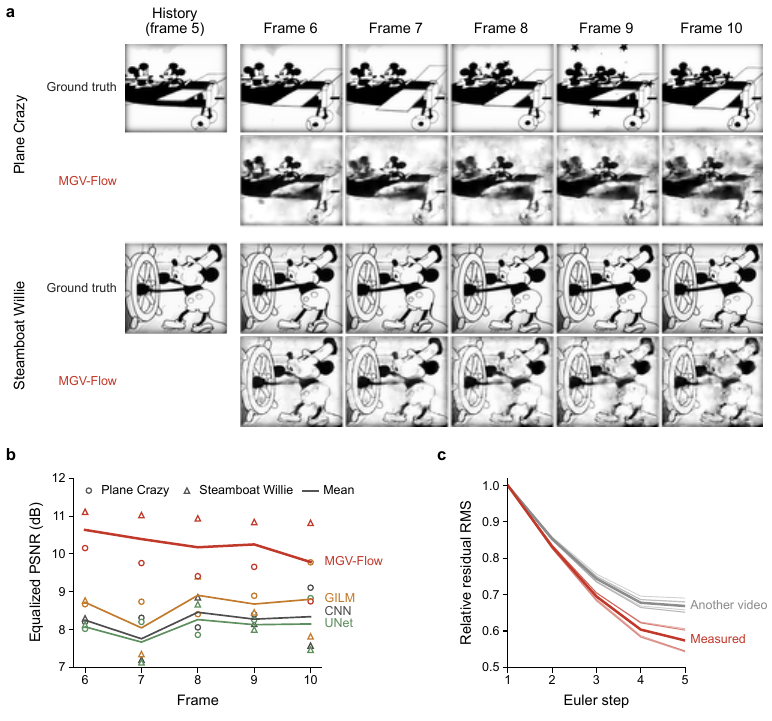}
\caption{
Physical single-pixel photography reconstruction with 10{,}000 measured bucket values per frame.
\textbf{a}, For each clip, the last frame of the ground-truth history, the ground-truth frames 6 to 10, and the frames that Measurement-Gradient-Conditioned Video Flow (MGV-Flow) predicts from the history and the measured bucket values of frames 6 to 10, with random seed 20261019 and $K=5$ Euler steps. Tiles are rectified and contrast-stretched between the 1st and 99th percentiles.
\textbf{b}, Equalized peak signal-to-noise ratio (PSNR) of each frame for MGV-Flow, averaged over three seeds, and for per-frame reconstruction by a convolutional neural network (CNN), UNet and GILM. Markers are the two clips and lines their mean.
\textbf{c}, Root mean square (RMS) of the measurement residual $r$ along the five Euler steps of MGV-Flow, normalized by its value at the first step, with the measured bucket values and with those of the other clip ($M=10{,}000$). Thin lines are the six runs and thick lines their means.
}
\label{fig:physical}
\end{figure*}

\begin{table*}[t]
\caption{Reconstruction of physical frames 6 to 10. MGV-Flow, Measurement-Gradient-Conditioned Video Flow, uses the ground-truth frames 1 to 5 as history and either the measured bucket values, the bucket values of the other clip, or no measurement input, with means over two clips and three seeds. CNN (convolutional neural network), UNet, and GILM reconstruct every frame independently from 10{,}000 bucket values, with means over the ten frames. $M$ is the number of bucket values per frame. Metrics use the rectified support, and Eq. denotes independent histogram equalization of estimate and reference. Time is per frame for the per-frame methods and per five-frame clip for MGV-Flow, and time and peak graphics processing unit (GPU) memory were measured on an RTX~3090. PSNR is the peak signal-to-noise ratio and SSIM the structural similarity.}\label{tab:physical}
\centering
\small
\setlength{\tabcolsep}{3pt}
\begin{tabular}{llrrrrrrr}
\toprule
Method & Bucket values & $M$ & PSNR & SSIM & Eq. PSNR & Eq. SSIM & Time & Memory \\
 & & & (dB) & & (dB) & & (s) & (GiB) \\
\midrule
CNN  & measured & 10{,}000 & 7.40 & 0.117 & 8.22 & 0.089 & 11.17 & 2.52 \\
UNet & measured & 10{,}000 & 9.03 & 0.191 & 8.06 & 0.085 & 18.28 & 2.55 \\
GILM & measured & 10{,}000 & 6.35 & 0.238 & 8.63 & 0.205 & 44.13 & 21.72 \\
\midrule
MGV-Flow & measured & 10{,}000 & \textbf{12.08} & 0.514 & \textbf{10.25} & 0.328 & 7.85 & 2.76 \\
MGV-Flow & measured & 1{,}000 & 12.06 & \textbf{0.518} & \textbf{10.25} & \textbf{0.329} & & \\
MGV-Flow & other clip & 10{,}000 & 11.94 & 0.509 & 10.09 & 0.325 & & \\
MGV-Flow & none & & 11.85 & 0.491 & 9.63 & 0.295 & & \\
\bottomrule
\end{tabular}
\end{table*}

MGV-Flow recovered recognizable frames from the physical measurements (Fig.~\ref{fig:physical}a and Table~\ref{tab:physical}). With the measured bucket values, its predictions reached an equalized PSNR of 10.25~dB and SSIM of 0.328, or 12.08~dB and 0.514 without equalization, whereas the best per-frame method, GILM, reached 8.63~dB and 0.205 on the same frames and CNN and UNet stayed near 8~dB, and on average over the two clips MGV-Flow exceeded every per-frame method on average over the two clips at each frame index, and in nine of the ten individual frames (Fig.~\ref{fig:physical}b). The raw and equalized rankings of the per-frame methods differ because the photometric mismatch affects each method differently. MGV-Flow produced all five frames in 7.85~s with 2.76~GiB of peak GPU memory, whereas GILM required 44.13~s and 21.72~GiB for each frame.

The controls show that the model reads the physical measurements, although most of the accuracy on this system comes from the accurate history. Replacing the measured bucket values with those of the other clip lowered the equalized PSNR in all six runs, by 0.16~dB on average and between 0.11 and 0.21~dB, and removing the measurement input lowered it in all six runs by 0.62~dB on average. Along the five Euler steps, the root mean square of the residual with respect to the measured bucket values fell to between 0.54 and 0.61 of its initial value, against 0.65 to 0.69 for the bucket values of the other clip (Fig.~\ref{fig:physical}c), so the trajectory moves the estimate toward agreement with the measurements of the target. With only the first 1{,}000 bucket values per frame, the rate used in simulation, the equalized PSNR was unchanged at 10.25~dB. A complete end-to-end system would also require reconstructing the history frames from the same measurements and radiometric calibration of the detector chain.

\section{Discussion}\label{sec3}

At 0.24\% of the pixel count, single-pixel measurements no longer form an image, yet they still test every estimate, and the residual image turns this test into a reconstruction signal. Evaluated in the latent space of a pretrained video decoder, it acquires spatial structure that decoders of the same architecture without learned weights do not produce, and joint reconstruction of a clip in that space is more accurate and cheaper than frame-wise image-latent reconstruction. Read by a flow model together with the preceding frames, it supports reconstruction of a clip in about a second rather than minutes, and the measurements add to the history mainly where motion leaves the future uncertain. Its direction, which depends on pattern family, grain size, and latent representation, ranks coding configurations by the quality of physics-only steps before that reconstruction is run.

Single-pixel photography systems have relied on per-frame reconstruction, deterministic or adaptive coding, and recurrent or physics-driven networks trained for a fixed acquisition \cite{edgar2015simultaneous,higham2018deep,lorente2020recurrent,xu2024hyperspectral,xu2026broadband}, and generative and latent priors have so far entered single-pixel and correlation imaging mainly for still images \cite{chen2025large,souza2025latent,song2024fspi}. Posterior-guidance methods, in turn, combine generative priors with arbitrary measurements by adding the measurement gradient with a prescribed weight at every sampling step \cite{chung2023dps,rout2023solving,kim2025flowdps,kwon2025solving}. The approach studied here keeps the physical gradient of correlation imaging as the measurement interface and lets a video model trained once decide how to combine it with temporal context, without a fixed acquisition network, a hand-tuned guidance weight or a per-clip optimization. Added with a fixed weight, the same direction degraded reconstruction with few steps and became competitive only with ten.

\begin{figure*}[!tp]
\centering
\includegraphics[width=0.62\textwidth]{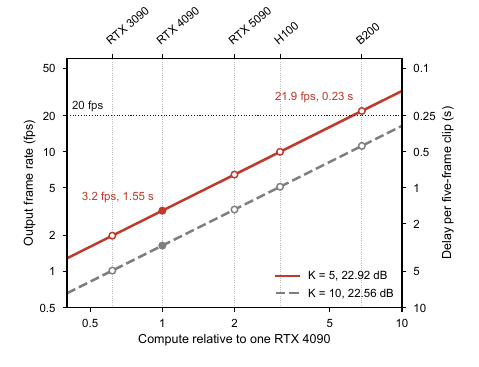}
\caption{
Imaging rate of sliding-window single-pixel photography versus compute. Filled markers are the median latencies measured on one RTX~4090 for $K=5$ and $K=10$ Euler steps at $M=1{,}000$, with the median peak signal-to-noise ratio (PSNR) in the legend. Open markers are theoretical values that assume latency inversely proportional to compute throughput, with the RTX~3090 scaled by its measured latency ratio to the RTX~4090 and the RTX~5090, H100 and B200 by nominal throughput ratios of 2.0, 3.1 and 6.8. The right axis gives the delay, the latency of one five-frame clip, and the dotted line marks 20~frames per second.
}
\label{fig:imaging-rate}
\end{figure*}

Because the detector delivers bucket values as a continuous stream, the reconstruction can run as a sliding window that reads the most recent $M=1{,}000$ bucket values per frame together with the five preceding reconstructed frames, so that the reconstruction delay is the latency of one five-frame clip. On a single RTX~4090 this latency was 1.55~s at $K=5$ and 3.04~s at $K=10$, which corresponds to sustained output rates of 3.2 and 1.6~frames per second; with ground-truth preceding frames the median PSNR was 22.92 and 22.56~dB, respectively (Fig.~\ref{fig:imaging-rate}). If the latency scales inversely with compute throughput, the measured RTX~3090 to RTX~4090 latency ratio gives about 2.0~frames per second on an RTX~3090, and theoretical throughput ratios place an RTX~5090 at 6.4, an H100 at 10.0, and a data-centre accelerator at 21.9~frames per second.

Within each frame the centered measurements constrain at most $M-1$ of the $HW$ pixel directions, so a large set of videos agrees with the measurements almost equally well. Per-clip optimization reduces the measurement error within that set without a preference among its members, and accordingly fitted the measurements far more closely than MGV-Flow but reconstructed worse video. A learned model can instead use the preceding frames and its training distribution to select a plausible member. Likewise, the decoder response is the measurement-derived direction reshaped by the local Jacobian of the pretrained decoder, and its agreement with the true error is an empirical finding under the stated protocol, not a guarantee that the unmeasured components are recovered.

Reading the measurements pays off in high-motion scenes, where the preceding frames predict the target least, and the average gain over a history-only model across all motion levels is small. At ten Euler steps, an explicit gradient correction with a tuned weight becomes competitive on cross-dataset scenes, where the larger step budget allows stronger measurement guidance. The simulations are noiseless and the flow model was trained on about 10K samples from one video collection, so performance under detector noise and with more diverse training data has not been characterised. Directional fidelity uses the ground-truth latent and therefore ranks coding configurations as an offline analysis; an adaptive analogue would require an estimate of the latent correction without access to the target. The physical experiment used accurate ground-truth history frames, and extending it to reconstructed histories requires addressing the residual photometric mismatch between the simulated and physical operators.

For any differentiable forward model and data-fidelity objective, the measurement discrepancy can be propagated to the reconstruction variables through the adjoint of the forward-model Jacobian, and for the linear single-pixel model considered here this operation reduces to the residual-weighted correlation with the coding patterns that defines the residual image. The same reading could therefore serve single-pixel photography at wavelengths where detector arrays are unavailable or costly, X-ray single-pixel tomography \cite{kingston2019ghosttomography}, and other computational imaging systems with a known forward model, provided that a generative model of the target class is available. When measurements are scarce, the present results suggest evaluating this feedback in a representation learned from the target class, reading it together with context that the measurements cannot supply, and choosing the measurements for the direction they give.

\section{Methods}\label{sec4}

\subsection{Measurement model and residual image}\label{sec4-measurement}

Each reconstructed grayscale clip contains $T=5$ frames $x_t\in[0,1]^{H\times W}$, with $H=480$ and $W=856$ in the target setting. For $M$ binary coding patterns $P_m\in\{0,1\}^{H\times W}$, the forward operator $A$ is
\begin{equation}
  (Ax)_{t,m}=\frac{\langle P_m,x_t\rangle}{HW},
  \qquad t=1,\ldots,T,\quad m=1,\ldots,M,
  \label{eq:forward}
\end{equation}
and the simulated bucket values are $y=Ax^\ast$ for the original ground-truth clip $x^\ast$. The frozen LeanVAE encoder $E$ and decoder $D$ map video clips and latents, with $z^\ast=E(x^\ast)$ and estimate $\hat x=D(z)$ at a latent $z$.

At a latent estimate $z$, we compute the measurement residual $r$ and remove its measurement mean independently in every frame:
\begin{equation}
  r=AD(z)-y,
  \qquad
  \tilde r_{t,m}=r_{t,m}-\frac{1}{M}\sum_{j=1}^{M}r_{t,j}.
  \label{eq:residual}
\end{equation}
For the centered quadratic measurement loss
\begin{equation}
  \mathcal L_{\mathrm c}=\frac{1}{2TM}\sum_{t,m}\tilde r_{t,m}^{\,2},
  \label{eq:centered-loss}
\end{equation}
the image-space residual image $g_x$ is the gradient with respect to the frame, obtained by backprojecting $\tilde r$ through the adjoint of $A$:
\begin{equation}
  (g_x)_t=\frac{1}{TMHW}\sum_{m=1}^{M}\tilde r_{t,m}P_m.
  \label{eq:gx}
\end{equation}
With this normalization, the equality is exact because $\sum_m\tilde r_{t,m}=0$.
It has the form of traditional correlation imaging, with the centered residual replacing the centered bucket signal. Linearity therefore gives
\begin{equation}
  g_x\ \propto\ \operatorname{TCI}[AD(z)]-\operatorname{TCI}[y].
  \label{eq:tgi-identity}
\end{equation}
One vector-Jacobian product through the frozen decoder lifts this signal into video latent space as the lifted residual image
\begin{equation}
  g_z=J_D(z)^{\mathsf T}g_x,
  \qquad J_D(z)=\frac{\partial D}{\partial z}.
  \label{eq:gz}
\end{equation}
The vectors $g_x$ and $g_z$ are ascent directions, and their negatives are the corresponding descent directions.
If $y=Ax^\ast+\epsilon$, then the residual also contains the noise term $-\epsilon$.
Constructing $r$, $g_x$, and $g_z$ at reconstruction time requires no target frames, whereas training, the directional-fidelity analysis, and PSNR or SSIM evaluation use reference frames.
The reconstruction-time calculation is detached before it enters the learned sampler and does not create a trainable backward path through $D$.

The image-space effect of a latent step along $-g_z$ is the decoder response $J_D(z)(-g_z)$, which we estimate by a central finite difference along the unit direction $u=-g_z/\lVert g_z\rVert_2$ with step $\delta=0.01\lVert z\rVert_2$,
\begin{equation*}
  J_D(z)(-g_z)\approx\lVert g_z\rVert_2\,\frac{D(z+\delta u)-D(z-\delta u)}{2\delta},
\end{equation*}
where $\delta=\epsilon\lVert z\rVert_2$ and $\epsilon$ was the smallest of the candidate values $0.01$, $0.02$ and $0.04$ that gave a nonzero output, which was $\epsilon=0.01$ in every case. No step-halving or Jacobian-vector-product comparison was run, and the reported Pearson correlations and band coherences do not depend on the amplitude of this estimate.
To test whether its structure originates in the learned weights, we repeated the calculation at the same latent state and with the same $g_x$ for control decoders of identical architecture.
Re-initialized controls redraw the parameters with the default initializer of each module, with Kaiming uniform initialization for the remaining multi-dimensional tensors, and weight-shuffled controls randomly permute the elements of each parameter tensor, which preserves the value distribution of every layer.
Both controls replace 99.9\% of the decoder parameter elements, leave unchanged the buffers that hold the fixed wavelet filters and four latent-bottleneck tensors, and use three random seeds each.
For a map $a$ and the true error $e=x^\ast-\hat x$ with two-dimensional discrete Fourier transform $\mathcal F$, complex conjugate $\overline{\,\cdot\,}$ and frequency index $\xi$, the spectral coherence in a radial frequency band $B$ is
\begin{equation}
  \kappa_B(a,e)=\frac{\operatorname{Re}\sum_{\xi\in B}(\mathcal F a)_\xi\,\overline{(\mathcal F e)_\xi}}{\bigl(\sum_{\xi\in B}\lvert(\mathcal F a)_\xi\rvert^{2}\sum_{\xi\in B}\lvert(\mathcal F e)_\xi\rvert^{2}\bigr)^{1/2}},
  \label{eq:coherence}
\end{equation}
computed per frame after mean removal and a separable Hann taper, in 25 equal radial bands from 0 to 0.5 cycles per pixel.
Multiplying $\mathcal F a$ by a positive real factor that is constant within $B$ leaves $\kappa_B$ unchanged, whereas gains that vary with frequency inside a band can change it.
The spectrum-matched control multiplies the mean-removed $-g_x$ in the Fourier domain, without a taper, by the real non-negative gain $(P_{\mathrm{resp}}/P_{g_x})^{1/2}$, where $P_{\mathrm{resp}}$ and $P_{g_x}$ are the radial power spectra of the decoder response and of $-g_x$ in 60 bins of width 1/120 cycles per pixel, so that its radial power spectrum matches that of the decoder response.
Coherences are aggregated over 15 window and frame samples with interquartile bands, after averaging each control over its three seeds.

\subsection{Video latent reconstruction baselines}

The scaling benchmark compared TCI, a pixel-space CNN, a pixel-space UNet, GILM with an image VAE latent, and VLI, which optimizes a video diffusion transformer (V-DiT) network in the joint LeanVAE video latent. TCI used streaming prefixes of the coding patterns. Each learned baseline optimized a newly initialized reconstruction network for every test frame or clip, and GILM and VLI also optimized their pretrained decoders, so VLI is a per-instance network optimization rather than latent-variable descent.

The benchmark used the first five frames of five test videos at $256\times144$, $512\times288$, $768\times432$, and $856\times480$ pixels, with an independently seeded bank of independent and identically distributed (IID) Bernoulli patterns for each resolution. All methods used the same nested prefixes for $M$ from 1,000 to 10,000 in increments of 1,000 and fitted $y=Ax^\ast$ of the original frames, which were converted to BT.601 grayscale in $[0,1]$.

Each learned method minimized its own standardized measurement loss, in which predicted and acquired bucket values are standardized by $\mathcal S$, centering and division by the standard deviation over $m$ within each frame. For the VLI network $f_\phi$ with parameters $\phi$, this per-frame single-pixel imaging (SPI) objective was
\begin{equation}
  \mathcal L_{\mathrm{SPI}}
  =\frac{1}{TM}\sum_{t=1}^{T}\sum_{m=1}^{M}
  \left[
  \mathcal S\!\left((AD(f_\phi))_{t,m}\right)
  -\mathcal S\!\left(y_{t,m}\right)
  \right]^2,
  \label{eq:lspi}
\end{equation}
and CNN, UNet and GILM used the same form with their own network output. The optimizer backpropagated this loss into the network parameters and, for GILM and VLI, into the trainable decoder. The resulting update signal is related to $g_x$ but differs from it, because standardization removes the affine scale of the prediction, so that predicted bucket values equal to a positive affine rescaling of $y$ have zero standardized error but a nonzero centered residual and the two losses have different zero sets. The formal caps were 300 updates for CNN, 2,000 for UNet, 500 for GILM, and 1,500 for VLI, and metrics were evaluated every 50 updates.

For the matched latency experiment, VLI used 1,500 AdamW \cite{loshchilov2019decoupled} updates with learning rates $10^{-3}$ for $f_\phi$ and $10^{-5}$ for the decoder, betas $(0.9,0.99)$, zero weight decay, and gradient-norm clipping at 1.0.

Floating-point operations (FLOPs) included the reconstruction network and physical operator, with one multiply plus one add counted as two FLOPs. Reported FLOPs were accumulated to the best observed checkpoint and divided by the number of reconstructed frames. Peak memory was the process-local maximum allocation above the pre-reconstruction baseline and included model weights, gradients, optimizer state, activations, the decoder, and the pattern minibatch.

\subsection{MGV-Flow}

MGV-Flow predicts a velocity field $v_\theta$ with network parameters $\theta$ in the LeanVAE video latent. The history encoding $h$ is the frozen encoding of the preceding five grayscale frames. Flow time follows the convention $\tau=0$ for Gaussian noise and $\tau=1$ for data, with state $z_\tau$. Starting from a seeded Gaussian latent $z_0$, $K$ explicit Euler steps use
\begin{equation}
  z_{\tau+1/K}=z_\tau+\frac{1}{K}
  v_\theta\!\left(z_\tau,h,d_{\mathrm{phy}},q,\tau\right),
  \qquad \tau=\frac{k}{K},\quad k=0,\ldots,K-1.
  \label{eq:euler}
\end{equation}

At every step, equation~(\ref{eq:gz}) is evaluated once at $z_\tau$ and converted to the physical direction, which is first normalized to unit root mean square (RMS) over all latent elements and then gated,
\begin{equation}
\begin{split}
  d_{\mathrm{phy}}
  &=-\frac{g_z}{\bigl(\operatorname{RMS}(g_z)^2+10^{-24}\bigr)^{1/2}}
  \operatorname{gate}(\tau),\\
  \operatorname{gate}(\tau)
  &=\operatorname{smoothstep}\!\left(\operatorname{clip}_{[0,1]}
  \frac{\tau-0.15}{0.85}\right),
  \qquad \operatorname{smoothstep}(c)=3c^2-2c^3,
  \label{eq:dphy}
\end{split}
\end{equation}
so that its RMS equals $\operatorname{gate}(\tau)$ and vanishes for $\tau\le0.15$. No gradient clipping is applied to this direction. The quality descriptor, with natural logarithms and RMS values taken over all $T\times M$ centered residuals or all latent elements, is
\begin{equation}
\begin{split}
  q=\bigg[&\log\frac{M}{1000},
  \log\max\left\{\mathcal L_{\mathrm c},10^{-6}\right\},
  \log\max\{\operatorname{RMS}(\tilde r),10^{-6}\},\\
  &\log\max\{\operatorname{RMS}(g_z),10^{-6}\},
  \operatorname{gate}(\tau)\bigg].
  \label{eq:q}
\end{split}
\end{equation}
It describes measurement count, residual level, gradient scale and flow time, and its second and third components are deterministically related through $\mathcal L_{\mathrm c}=\operatorname{RMS}(\tilde r)^2/2$ apart from the floors.

The LeanVAE latent of a five-frame $480\times856$ clip, whose grayscale frames are repeated over three channels and shifted by $-0.5$, has shape $16\times2\times60\times107$ after eightfold spatial and causal $4n+1$ temporal compression. The network projects $z_\tau$, $d_{\mathrm{phy}}$ and $h$ to 288 channels by $1\times1\times1$ convolutions and adds the projected $z_\tau$, the projected $d_{\mathrm{phy}}$ and the time mean of the projected history. Sinusoidal flow-time embedding followed by a multilayer perceptron and $q$ passed through layer normalization and a multilayer perceptron are summed into one conditioning vector, which modulates every residual block by feature-wise linear modulation and the cross-attention by adaptive layer normalization. The backbone is a three-dimensional convolutional U-Net with three residual blocks at 288 channels, stride-2 downsampling to 576 channels, three further blocks, one cross-attention layer with 8 heads and width 576 whose keys and values are the $2\times2$-pooled time-mean history, transposed-convolution upsampling with skip concatenation, three final blocks and a zero-initialized output, giving 107,515,898 parameters. It predicts the complete velocity, so $d_{\mathrm{phy}}$ is not hard-added to its output.

Training used the straight interpolation $z_\tau=(1-\tau)z_0+\tau z^\ast$ with $z_0\sim\mathcal N(0,I)$ and $\tau$ drawn uniformly from $[0,1]$. With the predicted clean latent $\hat z_1=z_\tau+(1-\tau)v_\theta$, the measurement-consistency loss
\begin{equation*}
  \mathcal L_{\mathrm{meas}}
  =\frac{1}{TM}\sum_{t,m}
  \left[\mathcal S\!\left((AD(\hat z_1))_{t,m}\right)-\mathcal S(y_{t,m})\right]^2
\end{equation*}
standardizes over all $T\times M$ values of the clip, by centering and division by the square root of the mean square plus $10^{-6}$, and reaches $\theta$ through the frozen decoder, which runs in BF16 and whose RGB output is converted to BT.601 grayscale. The objective was
\begin{equation}
  \mathcal L_{\mathrm{train}}
  =\operatorname{MSE}\!\left(v_\theta(z_\tau,h,d_{\mathrm{phy}},q,\tau),z^\ast-z_0\right)
  +0.1\,\mathcal L_{\mathrm{meas}},
  \label{eq:training-loss}
\end{equation}
where the mean squared error (MSE) is averaged over all elements, and $d_{\mathrm{phy}}$ and $q$ are computed from the current $z_\tau$ and the clip's own simulated $y$ with $M=1{,}000$ and detached, so that no gradient passes through their construction. Training used about 10K samples at $856\times480$ of the training split, drawn from 17 source videos, with all windows of five history and five future frames at stride 1, giving 9{,}338 windows per epoch and 93{,}380 training windows over ten epochs or 11{,}680 steps with a global batch size of 8 on two RTX 3090 GPUs, and the final checkpoint was used. AdamW used learning rate $10^{-4}$, betas $(0.9,0.95)$, weight decay $10^{-4}$, gradient-norm clipping at 1.0, BF16 weights and seed 20260914. LeanVAE was frozen, and only the velocity model was trainable.

\subsection{Comparison conditions}

History-FM is a separately trained network $v_\eta$ with parameters $\eta$, of the same architecture and history $h$, trained with seed 20260912 and the same protocol except that $d_{\mathrm{phy}}$ and $q$ were set to zero in training and sampling and $\mathcal L_{\mathrm{meas}}$ had weight zero. The FlowDPS-inspired explicit posterior-gradient baseline used the step-11,680 History-FM checkpoint and added the physical direction at every Euler step:
\begin{equation}
  z_{\tau+1/K}=z_\tau+\frac{1}{K}
  \left[v_\eta\!\left(z_\tau,h,\tau\right)
  +\gamma d_{\mathrm{phy}}\right],
  \qquad \gamma=1.0.
  \label{eq:flowdps}
\end{equation}
The preset guidance weight $\gamma$ was not tuned by $K$, by suite, or on the cross-dataset scenes. History-FM, FlowDPS-inspired, and MGV-Flow used the same Gaussian initial latent, Euler grid, pattern identifiers, LeanVAE, and measurement operator.

The zero control set $d_{\mathrm{phy}}$ and all five components of $q$ to zero at sampling without evaluating the physics. Unlike for History-FM, this is an out-of-distribution input for MGV-Flow, so the shuffled control is the stricter test. The shuffled control kept the target latent and history fixed but replaced $y$ with bucket values from a distinct donor video, with matched pattern identifiers, $M$, solver step, and evaluation seed, and recomputed the residual, direction, and descriptor from the donor bucket values. We define $\Delta_{\mathrm{zero}}$ and $\Delta_{\mathrm{shuf}}$ as the correct-conditioning PSNR minus the corresponding zero-conditioning or shuffled-conditioning PSNR.

The matched flow evaluation used seven videos, five predetermined windows per video, three seeds, $M=1{,}000$, $H=480$, $W=856$, and $K\in\{2,3,4,5,8,10\}$. This produced 105 matched cases for each Euler budget and condition.

\subsection{Directional fidelity and coding-pattern study}

Directional fidelity measures whether the lifted residual image points toward the true latent correction. With Gaussian noise $n\sim\mathcal N(0,I)$ scaled by the global standard deviation of $z^\ast$ and seeded by 20260814 plus the window index, the probe latent at probe time $\tau=0.5$ was
\begin{equation}
  z_{\mathrm{pr}}=(1-\tau)n+\tau z^\ast,
  \qquad \tau=0.5.
  \label{eq:probe}
\end{equation}
The directional fidelity $C$ was
\begin{equation}
  C=\cos\angle\!\left(-g_z(z_{\mathrm{pr}}),z^\ast-z_{\mathrm{pr}}\right).
  \label{eq:directional-fidelity}
\end{equation}
Each value required one residual image and one decoder vector-Jacobian product. Every coding-pattern configuration and every video autoencoder used the same noiseless bucket values $y=Ax^\ast$ of the original frames, each autoencoder used its own $z^\ast=E(x^\ast)$, and the same pattern prefix served for $C$ and the corresponding reconstruction trajectory.

The study compared block Bernoulli, Walsh, low-frequency Fourier, Haar, and phase-speckle patterns at grain sizes of 4, 16, and 32 pixels, with $M=1{,}000$ and $M=3{,}000$. Block Bernoulli patterns set half of the coarse cells to one, Walsh patterns are parity patterns on a power-of-two grid with the constant pattern skipped, complements in the second half and nearest upsampling, and low-frequency Fourier patterns are cosines at quadrature phases ordered by radius and thresholded at their medians. Haar patterns are separable wavelet products thresholded at zero with complements in the second half, and phase-speckle patterns are the squared magnitude of the inverse Fourier transform of a random phase, bilinearly upsampled and thresholded at the median. Pattern seed 20260814 was fixed. Four windows were used for calibration and eight disjoint windows were held out for analysis. The 32 pixel structured banks contain finite numbers of distinct patterns, 1,022 for Walsh, 808 for low-frequency Fourier, and 958 for Haar, and pattern indices wrap cyclically beyond these maxima. For the images of Fig.~\ref{fig:pattern-gradients}, we recomputed three configurations on one held-out window with the same code, seeds and probe latent, saved the image-space arrays, and confirmed that $C$ and the terminal PSNR reproduce the stored values exactly. There the decoder response, evaluated by the central finite difference of Section~\ref{sec4-measurement}, is compared with the decoded latent correction $D(z^\ast)-D(z_{\mathrm{pr}})$.

For the decoder-response diagnostic defined in Section~\ref{sec4-measurement}, we selected the displayed windows by the highest, median, and lowest ground-truth motion scores.

Starting from the same $z_{\mathrm{pr}}$, the physics-only trajectory took ten updates
\begin{equation}
  z\leftarrow z-0.05\operatorname{RMS}(z_{\mathrm{pr}})
  \frac{g_z}{\max\{\operatorname{RMS}(g_z),10^{-12}\}}.
  \label{eq:pure-physics}
\end{equation}
The decoder output was not clamped during the updates, and the terminal decoded video was clamped to $[0,1]$ only for PSNR and SSIM evaluation against the original frames. The primary statistic was the Spearman correlation $\rho_{\mathrm s}$ across the 30 configuration-level held-out means, with a 95\% percentile interval from 10,000 resamples of the eight held-out windows with replacement and seed 20261008.

\subsection{Metrics and timing}

PSNR and SSIM were computed independently on each grayscale future frame in $[0,1]$, then averaged over $T=5$. PSNR used data range 1.0. SSIM used an $11\times11$ Gaussian window with standard deviation 1.5, data range 1.0, and constants $0.01^2$ and $0.03^2$.
In the scaling benchmark, Table~\ref{tab:quality-regime} and Fig.~\ref{fig:latent-benchmark}a use each run's best observed PSNR checkpoint with its paired SSIM, averaged over four resolutions and five videos, whereas Fig.~\ref{fig:latent-benchmark}b and the per-resolution comparison use the PSNR at the final update of each run.

Measurement NMSE centers the prediction and target over $m$ separately for each frame. With the centered bucket values $\tilde y_{t,m}=y_{t,m}-M^{-1}\sum_j y_{t,j}$, it is
\begin{equation}
  \mathrm{NMSE}
  =\frac{\sum_{t,m}\tilde r_{t,m}^{\,2}}
  {\max\left\{\sum_{t,m}\tilde y_{t,m}^{\,2},10^{-12}\right\}}.
  \label{eq:nmse}
\end{equation}
The sums cover all five future frames and all measurements. This evaluation metric uses centering but no variance standardization.

For an iterative reconstruction whose PSNR sequence is smoothed by a three-point median, with initial value $\mathrm{PSNR}_0$ and observed ceiling $\mathrm{PSNR}_{\max}$, the convergence time is the first evaluated time satisfying
\begin{equation}
  \mathrm{PSNR}\ge\mathrm{PSNR}_0+0.95\left(\mathrm{PSNR}_{\max}-\mathrm{PSNR}_0\right).
  \label{eq:t95}
\end{equation}
The baseline timing pass used one fixed test video's first five frames at $856\times480$ and $M=10{,}000$, recorded three repeats on each exclusive RTX 3090 and RTX 4090 including one warm-up update, and reports convergence times as medians over these repeats. Because the convergence time uses each method's own observed ceiling, it must be interpreted together with reconstruction quality.

The matched latency benchmark at $M=1{,}000$ started with in-memory reconstruction inputs. It included required initialization or history encoding and final pixel reconstruction. It excluded disk input and output, video decoding, checkpoint loading, synthetic measurement generation, metric computation, result writing, and paused trajectory diagnostics. The delivery records three repeats and does not document a warm-up procedure. Within every paired case, all methods ran on the same exclusive RTX 4090. Two physical cards were used across the full delivery. Latency and reconstruction metrics first took the median over repeats within each case and then the median over 21 paired cases. Iterative trajectories used the same two-stage median at each logged update.

The ground-truth motion score was the mean squared difference between the last history frame and the five future frames. It was computed on the original $856\times480$ colour (RGB) frames. Within each clip, ranks 1 and 2 formed the high-motion group, rank 3 the middle group, and ranks 4 and 5 the low-motion group. Reported associations used the Spearman correlation $\rho_{\mathrm s}$. The coding-pattern confidence interval followed the window bootstrap specified above.

\subsection{Physical single-pixel photography protocol}\label{sec4-physical}

The physical system uses structured illumination (Fig.~\ref{fig:optical-setup}). A laser beam, conditioned by two lenses and a beam expander and folded by a mirror, illuminates a digital micromirror device (DMD) that displays the binary patterns, two lenses project the patterns onto the scene, and a fifth lens collects the light from the scene onto the single-pixel detector. Two ten-frame excerpts of the animated films Plane Crazy and Steamboat Willie were acquired frame by frame with the same ordered bank of 10{,}000 binary speckle patterns used in simulation. The acquisition software normalizes every bucket value between black and all-white references, and we inverted the measured detector response with a separately acquired lookup table. Because the operator of equation~(\ref{eq:forward}) uses absolute intensity whereas the instrument stores ratios, the ratios of frames 6 to 10 were multiplied by the mean intensity of the five history frames, so that no future frame entered the inputs or this scaling.

The patterns illuminate a central quadrilateral of the $856\times480$ reconstruction grid, into which the history frames were warped. For evaluation, every estimate and reference was rectified to a common $408\times360$ rectangle, on which we report PSNR and SSIM before and after independent histogram equalization of estimate and reference.

CNN, UNet, and GILM reconstructed every frame independently from its measured bucket values, with 300 updates for CNN and UNet and 100 epochs for GILM, keeping the checkpoint with the lowest measurement loss. MGV-Flow predicted frames 6 to 10 with the ground-truth history and three seeds, from the measured bucket values with all 10{,}000 or the first 1{,}000 patterns, from the bucket values of the other clip scaled in the same way, or without measurement input. Time and peak memory were recorded on one exclusively used RTX~3090.

\bmhead{Acknowledgements}

We thank the authors of LeanVAE \cite{cheng2025leanvae}, whose pretrained video autoencoder provides the latent space of MGV-Flow, and of VidTok \cite{tang2024vidtok}, Wan \cite{wan2025wan} and HunyuanVideo \cite{kong2024hunyuanvideo}, whose autoencoders we use in the directional-fidelity analysis. The FlowDPS-inspired baseline follows \cite{kim2025flowdps}. Test sequences come from the Xiph.org Derf collection and Something-Something V2 \cite{goyal2017something}. Codex (OpenAI) and Claude Code (Anthropic) assisted with the discussion of ideas and with language editing, and the authors take full responsibility for the content.

\section*{Data availability}

\begin{itemize}
\item LeanVAE pretrained weights (video decoder): \url{https://github.com/westlake-repl/LeanVAE}
\item VidTok video tokenizer: \url{https://github.com/microsoft/VidTok}
\item Wan video autoencoder: \url{https://github.com/Wan-Video/Wan2.1}
\item HunyuanVideo autoencoder: \url{https://github.com/Tencent-Hunyuan/HunyuanVideo}
\item Xiph.org Derf test sequences: \url{https://media.xiph.org/video/derf/}
\item Something-Something V2 \cite{goyal2017something}: \url{https://www.qualcomm.com/developer/software/something-something-v-2-dataset}
\end{itemize}

\section*{Code availability}

Code is available at \url{https://github.com/cavedweller509/MGV-Flow}.

\section*{Author contributions}

H.A. conceived the study and wrote the manuscript. S.A. assisted with the experiments and data processing. T.T. assisted with building the physical optical setup. Z.S. led the work, including supervision and review of the manuscript. All authors approved the final manuscript.


\bibliography{sn-bibliography}


\end{document}